\documentclass[
 floatfix,
 aps,
 prl,
 amsmath,
 amssymb,
 superscriptaddress,
 longbibliography,
 reprint
]{revtex4-2}

\usepackage{afterpage}          
\usepackage{bm}                 
\usepackage{booktabs}           %
\usepackage{braket}             %
\usepackage{dcolumn}            
\usepackage{etoolbox}           %
\usepackage{flafter}            %
\usepackage[T1]{fontenc}        %
\usepackage{graphicx}           
\usepackage{hhline}             %
\usepackage[utf8]{inputenc}     %
\usepackage{mathtools}          %
\usepackage{microtype}          %
\usepackage{natbib}             %
\usepackage{relsize}            %
\usepackage{siunitx}            %
\usepackage{titlesec}           %
\usepackage{xcolor}             %
\usepackage{newtxmath}          %
\usepackage{newtxtext}
\usepackage{fix-cm}
\usepackage[hypertexnames=false]{hyperref}   
\usepackage{multirow}
\usepackage{tabularx}

\makeatletter
\def\@email#1#2{%
 \endgroup
 \patchcmd{\titleblock@produce}
  {\frontmatter@RRAPformat}
  {\frontmatter@RRAPformat{\produce@RRAP{*#1\href{mailto:#2}{#2}}}\frontmatter@RRAPformat}
  {}{}
}
\makeatother

\newcommand{\tempmd}{18.6}          
\newcommand{\tuncertmd}{1.5}          

\newcommand{\templd}{17.1}          
\newcommand{\tuncertld}{1.5}          

\newcommand{\densmd}{25}          
\newcommand{\densld}{24}          
\newcommand{\densuncert}{3}          
\newcommand{\ionmd}{6.0}            
\newcommand{\ionuncertmd}{0.1}
\newcommand{\ionld}{5.4}            
\newcommand{\ionuncertld}{0.2}

\newcommand{\dfttemp}{23}
\newcommand{\dftdens}{28}

\newcommand{\dfttempmd}{27}
\newcommand{\dftdensmd}{28}

\DeclareSIUnit\hartree{Ha}
\DeclareSIUnit\bohr{Bohr}

\begin{document}

\preprint{APS/123-QED}

\title{Informing spectral models for dense plasmas with K-edge absorption measurements of warm dense copper}

\author{T. Cordova}
  \email[contact email:]{tcordova@ucsd.edu}
  \affiliation{Lawrence Livermore National Laboratory, Livermore, CA, USA}
  \affiliation{Department of Mechanical \& Aerospace Engineering, University of California San Diego, San Diego, CA, USA}
\author{E. V. Marley}
\author{D. A. Chin}
  \affiliation{Laboratory for Laser Energetics, University of Rochester, Rochester, NY, USA}
\author{R. A. London}
  \affiliation{Lawrence Livermore National Laboratory, Livermore, CA, USA}
\author{S. B. Hansen}
  \affiliation{Sandia National Laboratories, Albuquerque, NM, USA}
  \author{S. M. Vinko}
  \affiliation{Department of Physics, Clarendon Laboratory, University of Oxford, Oxford, UK}
\author{J. E. Pask}
\author{H. A. Scott}
  \affiliation{Lawrence Livermore National Laboratory, Livermore, CA, USA}
\author{H. P. Le}
\author{D. {\AA}berg}
\author{M. K. G. Kruse}
\author{T. D{\"o}ppner}
  \affiliation{Lawrence Livermore National Laboratory, Livermore, CA, USA}
\author{F. N. Beg}
  \affiliation{Department of Mechanical \& Aerospace Engineering, University of California San Diego, San Diego, CA, USA}
\author{J. Emig}
  \affiliation{Lawrence Livermore National Laboratory, Livermore, CA, USA}
\author{P. M. Nilson}
  \affiliation{Laboratory for Laser Energetics, University of Rochester, Rochester, NY, USA}
\author{P. Sterne}
  \affiliation{Lawrence Livermore National Laboratory, Livermore, CA, USA}
\author{M. J. MacDonald}
  \affiliation{Lawrence Livermore National Laboratory, Livermore, CA, USA}

\date{\today}

\begin{abstract}
Warm dense matter remains a challenging regime to characterize experimentally and to model with predictive accuracy. Recent experimental platforms have been developed to generate, characterize, and diagnose uniform warm dense matter, enabling detailed comparisons with models. Here, we present experiments conducted at the OMEGA laser facility that compress and heat a buried layer target to warm dense matter conditions, where the targets are heated to temperatures of approximately \qty{20}{\electronvolt} and compressed to densities of \qty{25}{\gram\per\cubic\centi\metre}. We probe the warm dense plasma using x-ray absorption spectroscopy, using the K-edge and bound-bound absorption features to constrain the temperature and charge state distribution of the plasma. We compare these measurements with two types of models: collisional-radiative models with detailed electronic structure and ad-hoc density effects, and a multi-ion model based on density functional theory in combination with excited-state projector augmented-wave potentials. Neither approach fully reproduces the observed data, We show that the broad structure and position of the K-edge region can be modeled using density functional theory in combination with excited-state projector augmented-wave potentials. The density functional theory results are contrasted with a collisional-radiative model approach that incorporates ad-hoc density effects, which show incomplete agreement with the experimental observations, highlighting a need for improved density-dependent atomic modeling in warm dense plasmas.
\end{abstract}

\maketitle
\section{Introduction}\label{sec:introduction}
Understanding the behavior of atoms in dense, partially ionized plasmas is a critical element of understanding high energy density plasma systems such as inertial confinement fusion\cite{zylstraBurningPlasmaAchieved2022} and astrophysical bodies\cite{osterbrockInternalStructureRed1953,boothLaboratoryMeasurementsResistivity2015}. Regions of the interiors of exoplanets and stars consist of warm dense matter (WDM), characterized by moderate (several eV) temperatures, appreciable coupling, and a degenerate electron population that is strongly influenced by the surrounding dense plasma environment. Ionization in this environment can be enhanced due to density effects such as ionization potential depression (IPD)\cite{ciricostaDirectMeasurementsIonization2012} and pressure ionization or impeded due to Pauli blocking. These combined effects lead to shifts in absorption edges and changes in charge state distributions. Additionally, density-induced changes in the relative binding energies of electrons in different orbitals can lead to more subtle observable effects such as plasma polarization shifts in bound-bound features. In general, models developed for the limiting regimes of ideal plasmas and condensed matter cannot be directly extended to WDM. Due to these complexities, considerable work has been done to develop theoretical models and experimental platforms for WDM to investigate x-ray opacities, the equation-of-state, collisional plasma dynamics, and energy transport properties\cite{huProbingAtomicPhysics2022,hansenSelfconsistentDetailedOpacities2023,holstThermophysicalPropertiesWarm2008, vinkoCreationDiagnosisSoliddensity2012, hansenTemperatureDeterminationUsing2005, lutgertPlatformProbingRadiation2022,bailly-grandvauxLaserPulselengthDependent2024}.

Due to the high densities involved, established diagnostics based on visible or ultraviolet radiation cannot penetrate the plasma, making x-rays necessary to diagnose properties of WDM. Due to the limited x-ray self-emission of WDM plasmas, an external x-ray source is required to achieve sufficient signal quality for precise measurements. Experiments employing diagnostics such as x-ray radiography\cite{macdonaldCollidingPlanarShocks2023}, x-ray Thomson scattering\cite{bellenbaumModelfreeTemperatureDiagnostics2025}, and x-ray absorption spectroscopy (XAS)\cite{benuzzi-mounaixElectronicStructureInvestigation2011} have successfully measured and inferred temperature, density, and ionization in warm dense plasmas. However, accurate measurements remain sparse due to the experimental challenges of generating well-characterized WDM in the laboratory\cite{vorbergerRoadmapWarmDense2026}. WDM experiments can include significant temporal and spatial gradients that complicate analysis of the experimental observables\cite{macdonaldQuantifyingElectronTemperature2022}. 

\begin{figure*}
  \begin{center}
    \includegraphics[width=0.66\linewidth]{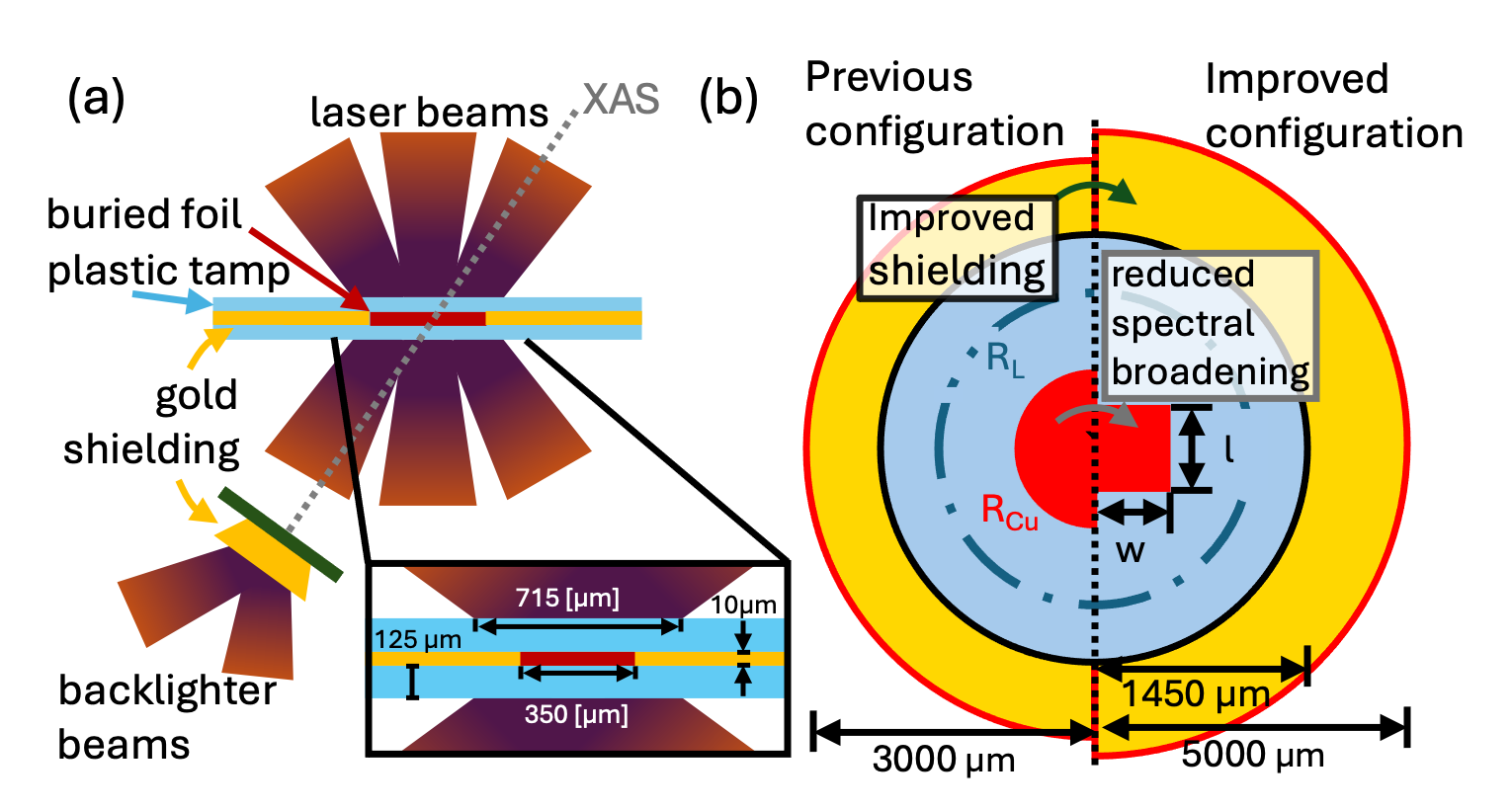}
    \includegraphics[width=0.33\linewidth]{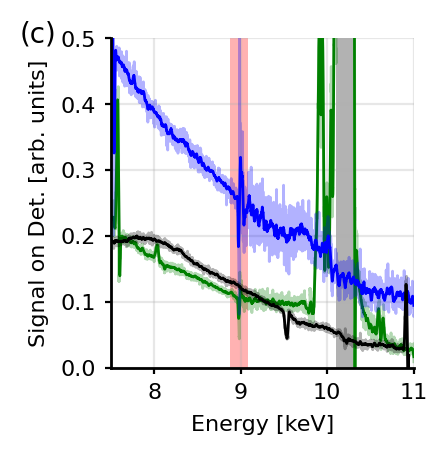}
    \caption{(a) Schematic of the experimental platform consisting of a buried layer tamped by CH and surrounded by an Au foil x-ray shield. A Ti or Ge backlighter provides x-rays timed to probe the uniform plasma created by the symmetric shocks, with either a \qty{500}{\pico\second} or \qty{300}{\pico\second} duration. (b) Top view of the target illustrating improvements to shielding, further extending the Au foil, and buried foil size, changed from $R_{Cu}=$\qty{500}{\micro\metre} disk to $w\times l = 200\times350$ \unit{\square\micro\metre} square, to reduce spatial gradients and spectral broadening along the line of sight. The laser spot size is set to overfill the buried layer size, $R_{L}=$\qty{715}{\micro\metre}. (c) Measured signal on the detector for the three backlighter types used in experiments, the Ge backlighters of previous configurations (green), the Ti backlighter (blue) yielded a $140$\% improvement at \qty{9}{\kilo\electronvolt}, and the Ti backlighter with consideration of the target geometry changes (black) resulted in a $14$\% signal improvement. The signals are filtered with a Savitzky-Golay filter for comparison purposes, and the vertical shaded area indicates the removal of a filter edge, Cu (red) and W (black).}
    \label{fig:1:omega_platform}
  \end{center}
\end{figure*}

In previous work, we developed a planar platform at the OMEGA laser facility in which we generated uniform warm dense copper plasmas and measured XAS\cite{cordovaIonizationTemperatureMeasurements2026} of the K-edge region. These experiments were designed to address some experimental challenges in generating a uniform volume of WDM and inferring temperature and charge state distribution (CSD) with the XAS. By varying the laser drive intensity, we observed clear sensitivity of the K-edge and bound-bound absorption to changes in temperature and ionization.

In this work, we present spectroscopic measurements of copper obtained using an improved platform at OMEGA, together with a detailed evaluation of modeling in this regime. Modeling in WDM conditions requires careful consideration of detailed atomic data, density effects, and excited state configurations. This study highlights how those details affect the predictive capability of different models, and that targeted improvements can bring better agreement with measurements. This experimental investigation was enabled by improvements to target design, timing control, and backlighter geometry. The experimentally inferred plasma conditions provide constraints for detailed comparisons with both collisional-radiative (CR) and density functional theory (DFT) based spectral models. Both modeling approaches capture the general characteristics of the observed spectra, such as the emergence of 3p vacancies and a softening of the absorption edge with increasing temperature. We find that the CR models provide a more complete picture of broadening due to multiple occupied electronic configurations while the DFT treatment provides a better match to the observed separation between bound-bound and bound-free spectral features. 

\section{Experimental setup and resulting data}\label{sec:experiment}
The experimental platform developed at the OMEGA laser facility generates WDM by compressing and heating a buried copper foil with symmetric laser-driven shocks. A time-delayed germanium backlighter probes the resulting WDM volume to measure XAS through the target. 
In our previous work\cite{cordovaIonizationTemperatureMeasurements2026}, we developed a simplified model to infer plasma conditions from the XAS measurements in combination with hydrodynamic simulations. Follow-up experiments introduced several improvements to enhance the XAS measurements. These improvements included additional shielding between the target and spectrometer, modification of the buried layer foil shape, changes to the backlighter duration and spot size, as well as employing titanium as the backlighter material.

\begin{figure*}
  \begin{center}
    \includegraphics[width=\linewidth]{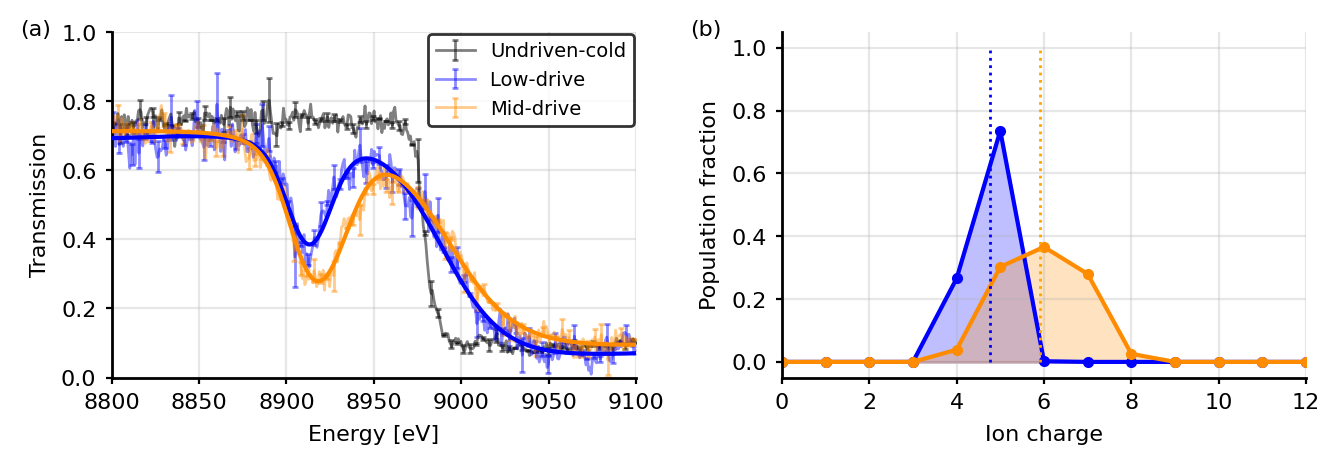}
    \caption{(a) Experimental XAS data of the low-drive (\qty{3}{\nano\second}) and mid-drive (\qty{2}{\nano\second}) configurations compared to the undriven-cold K-edge measurement, blue, orange, and black points with error bars. The model fit, in solid blue (low-drive) and orange (mid-drive) to the K-edge infers a temperature of \qty{\templd}{\electronvolt} and from the bound-bound absorption a $\overline{Z}=\ionld$ in the low-drive. We find an inferred temperature of \qty{\tempmd}{\electronvolt} and $\overline{Z}=\ionmd$ for the mid-drive. (b) The resulting fitted charge state distributions for the low- and mid-drive conditions.}
    \label{fig:2:exp_data}
  \end{center}
\end{figure*}

The target configuration consists of a planar \qty{10}{\micro\metre} thick rectangular copper foil (\qty{200}{\micro\metre} \texttimes\ \qty{350}{\micro\metre}) with a large gold washer of equal thickness surrounding it, acting as an x-ray shield to limit the absorption measurement to only the copper layer. This layer is placed between two equal \qty{125}{\micro\metre} thick plastic (CH) ablators. The size reduction, as compared to the previous \qty{500}{\micro\metre} disks, reduces source broadening effects on the measured XAS with the \qty{200}{\micro\metre} in the meridional plane.

The backlighter is a \qty{5}{\micro\metre} thick square titanium foil that provides improved continuum x-rays over the copper K-edge region around \qty{9}{\kilo\electronvolt}. XAS is collected through the target onto an image plate, effectively time gated to the backlighter probe duration.

The EFX spectrometer was employed for the XAS measurement\cite{chinHighresolutionXraySpectrometer2023}. EFX is a well-characterized spectrometer with a spectral range of \qtyrange[range-units=single]{6}{11}{\kilo\electronvolt}, allowing for multiple filter edges to constrain the dispersion relation for each shot. In previous work, an analytical estimate of the broadening was shown to agree well with the measured resolution of the undriven K-edge\cite{cordovaIonizationTemperatureMeasurements2026}. 
Since the spectral resolution in those previous experiments was primarily limited by the source size broadening, the reduced target size in the present platform directly improved the measured resolution. A diagram of the target package is shown in Figure \ref{fig:1:omega_platform}(a) and changes relative to the previous configuration are shown in Figure \ref{fig:1:omega_platform}(b). 

Additionally, the backlighter spot size was reduced from \qty{140}{\micro\metre} diameter to \qty{100}{\micro\metre} to further reduce source size broadening in the XAS measurements. A shorter \qty{300}{\pico\second} timing window reduced temporal gradients during the measurement window. The shorter pulse length configuration also reduced the total energy on target for the backlighter from approximately \qty{1200}{\joule} to \qty{700}{\joule}. The reduced spot size, combined with the smaller buried layer target, achieved an improved spectral resolution of \qty{7.9}{\electronvolt} at \qty{9}{\kilo\electronvolt} ($E / dE = 1140$); a $23$\% improvement over the previous experiment.

In comparison to previously used germanium backlighters, the switch to a titanium backlighter yields a $140$\% increase in signal at the copper K-edge\cite{doFoilBacklighterDevelopment2020}. Although the combined changes to the target reduce the overall photon flux to the spectrometer by 63\%, and the shorter duration further limits photon signal, the increased conversion efficiency of the titanium compensates for photons lost due to these changes. The Ti backlighter, combined with the target improvements, resulted in a $14$\% increase in signal at the copper K-edge. Figure \ref{fig:1:omega_platform}(c) shows a comparison of the measured signal obtained with the different measured backlighter sources Ge, Ti, and Ti with experimental configuration improvements used with this platform. 

\begin{table}[ht]
  \caption{Summarized table of simulated densities and temperatures form radiation-hydrodynamic code HYDRA for the low- and mid-drive. Included are the inferred values from the functional fit to K-edge and bound-bound transitions.}
  \label{tab:expvalues}
  \centering
  \renewcommand{\arraystretch}{1.25}
  \begin{tabularx}{\linewidth}{|l|c|c|}
    \hline
    \ & \ \ \ \ Low-drive\ \ \ \ \ &\ \ \ \ \ Mid-drive\ \ \ \ \\
    \hline
    Sim. Density [\unit{\gram\per\cubic\centi\metre}]
        & 23 -- 26 & 20 -- 28 \\
    Sim. Temperature [\unit{\electronvolt}]
        & 19 -- 22 & 27 -- 34 \\
    K-edge inferred temperature [\unit{\electronvolt}] & $\templd\pm\tuncertld$ & $\tempmd\pm\tuncertmd$\\ 
    Inferred Ionization $\overline{Z}$ & $\ionld\pm\ionuncertld$ & $\ionmd\pm\ionuncertmd$\\
    \hline
  \end{tabularx}
\end{table}

A simple model fit to the K-edge and bound-bound resonance absorption features provides an inferred temperature and charge state distribution from each XAS measurement. Previous work has shown that when the electron temperature is comparable to the Fermi energy of the plasma, the K-edge slope becomes sensitive to temperature through the Fermi-Dirac occupation of the electrons around the continuum edge \cite{dorchiesXrayAbsorptionEdge2015,hansenFluorescenceAbsorptionSpectroscopy2018}. This allows the temperature to be inferred directly from the edge slope. As temperature increases and multiple charge states are occupied, configurational broadening can also broaden the edge, making the simple estimate from the K-edge slope a lower bound.

Figure \ref{fig:2:exp_data} shows results for two experimental configurations obtained using the improved platform. These configurations are denoted as the low-drive and mid-drive corresponding to a \qty{3}{\nano\second} laser pulse length and a \qty{2}{\nano\second} laser pulse length, respectively. The on-target laser intensities for each configuration were $2\times10^{14}$ and $4\times10^{14}$ \unit{\watt\per\square\centi\metre} for the low- and mid-drive, respectively. Hydrodynamic simulations of the experiment incorporate custom laser-power multipliers constrained by dedicated VISAR measurements from the previous campaign\cite{cordovaIonizationTemperatureMeasurements2026}. 

Post-shot hydrodynamic simulations indicate both driven configurations achieved compressions to densities near \qty{25}{\gram\per\cubic\centi\metre} and that the mid-drive reached slightly higher temperatures. The changing density and temperature over the simulated probe durations provides the basis for creating an averaged fit with a simple functional model below, and for the CR models later in the discussion. Table \ref{tab:expvalues} provides a summary of the low- and mid-drive conditions extracted from the HYDRA simulations.

This result is consistent with the trends in the measured XAS, with the mid-drive data showing features consistent with higher temperature, that is, a deeper and broader bound-bound absorption feature and a shallower K-edge.  

The functional fits to the K-edge and bound-bound absorption feature developed in \cite{cordovaIonizationTemperatureMeasurements2026} provide useful constraints on the experimental plasma conditions. From the functional fit to the K-edge region, we infer a temperature of $\templd\pm\tuncertld$ \unit{\electronvolt} and an average ionization of $\overline{Z}=\ionld\pm\ionuncertld$ for the low-drive configuration. For the mid-drive configuration, we infer a temperature of $\tempmd\pm\tuncertmd$ \unit{\electronvolt} and an average ionization of $\overline{Z}=\ionmd\pm\ionuncertmd$. We estimate, via our hydrodynamic simulations, an averaged density of $\densld\pm\densuncert$ \unit{\gram\per\cubic\centi\metre} and $\densmd\pm\densuncert$ \unit{\gram\per\cubic\centi\metre} for the low- and mid-drives, respectively. The improved experimental fidelity enables tighter constraints on the inferred plasma conditions, with the inferred temperature acting as a lower bound. This allows for stringent comparison with CR and DFT-based spectral models in this regime where the dense plasma environment strongly influences the ionization balance as well as structure and position of the K-edge.

\section{Modeling x-ray absorption}\label{sec:modeling}
This work focuses on two modeling approaches that describe the electronic structure of ions in warm dense conditions and make predictions that can be compared with our experimental data. The treatment of density effects in each model directly influences the resulting predictions. 

The first approach utilizes CR models; here, we use both Cretin\cite{scottCretinRadiativeTransfer2001} and SCRAM. These models represent the electronic structure of atoms and ions using extensive sets of real (integer-occupied) electronic configurations that, depending on the ion and level of detail, can reach thousands of electronic states per modeled ion. Transitions among these states give rise to a set of coupled collisional-radiative rates that can be solved for level populations and used to predict detailed emission and absorption spectra. Here, Cretin uses tabulated data calculated with the Flexible atomic code (FAC)\cite{guFlexibleAtomicCode2008}. SCRAM uses FAC data for a subset of detailed states and transitions and supplements these with screened hydrogenic data to ensure model completeness. 

Both Cretin and SCRAM incorporate density effects, such as IPD and PPS, through ad-hoc modifications on top of the isolated atomic data. The IPD model employed in Cretin can be either the Stewart-Pyatt (SP) model\cite{stewartLoweringIonizationPotentials1966}, which interpolates between the low-density Debye limit and the high-density ion-sphere (IS) limit, or the Ecker-Kr{\"o}ll (EK) model approximation\cite{eckerLoweringIonizationEnergy1963}. SCRAM uses only IS or EK models. In both Cretin and SCRAM, PPS is approximated by the formulas for hydrogen-like ions by Nguyen et al\cite{nguyenAtomicStructurePolarization1986}, the gradual disappearance of bound states with density is achieved by reduction of statistical weights, and Fermi-Dirac statistics are used in modeling the photoionization edge and rates. Here, the major differences between Cretin and SCRAM lie in the details of their electronic level structure and their specific implementations of density effects (see \cite{jiangMeasurementsPressureinduced$Kensuremathbeta$2020} for additional details).  

\begin{figure*}
    \includegraphics[width=0.48\linewidth]{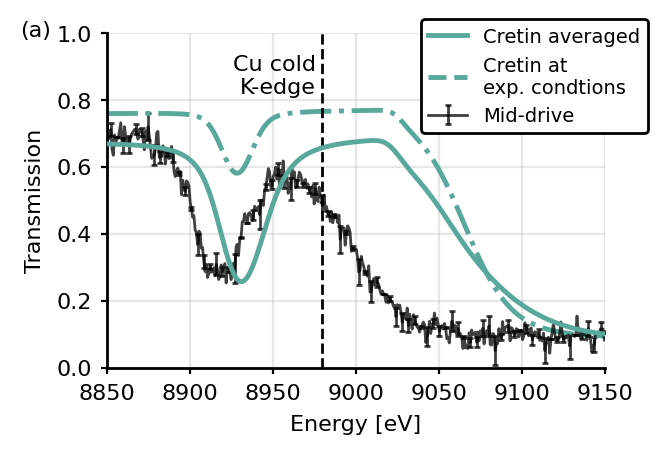}
    \includegraphics[width=0.48\linewidth]{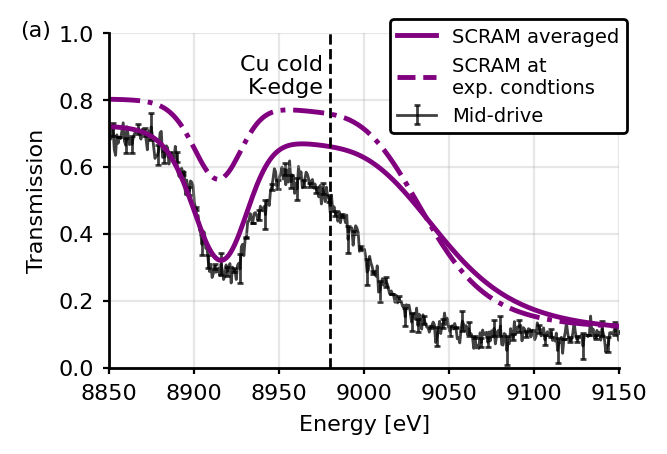}
    \includegraphics[width=0.55\linewidth]{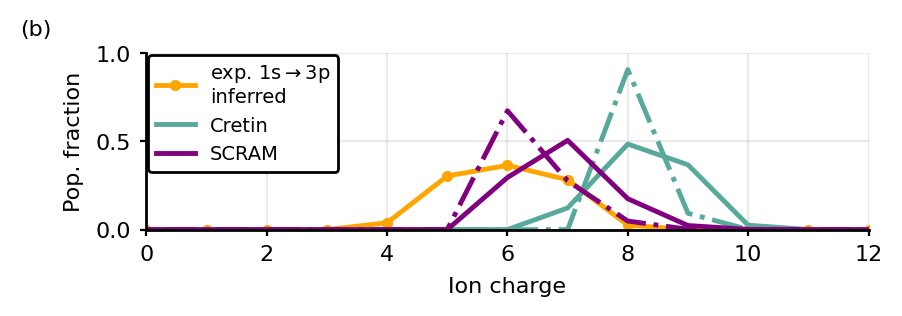}
    \caption{(a) Collisional-radiative code Cretin modeled at two conditions: At averaged conditions  of \qty{24}{\gram\per\cubic\centi\metre} and \qty{23}{\electronvolt} (solid). Single condition fit (dashed) at the inferred values from simulation for the density and experiment for the temperature of \qty{\densmd}{\gram\per\cubic\centi\metre} and \qty[round-mode=figures, round-precision=2]{\tempmd}{\electronvolt}, respectively. Comparison with the experimental XAS mid-drive is shown (black). (b) SCRAM modeled in the same manner, (solid) at averaged conditions and (dashed) at single experimental condition. (c) Comparison of the CSD of the Cretin and SCRAM, as compared with the experimentally inferred CSD.}
    \label{fig:model_crm}
\end{figure*}

The second modeling approach uses the finite-temperature DFT code ABINIT\cite{gonzeABINITFirstprinciplesApproach2009} to model electronic structure. We employed projector-augmented wave (PAW) potentials generated using ATOMPAW to approximate the copper ion cores\cite{holzwarthProjectorAugmentedWave2001}. For efficiency the 1s, 2s, 2p orbitals were treated with the frozen-core approximation, this is suitable as temperatures are too low to thermally ionize these orbitals. For the exchange-correlation functional, we have used the  Perdew-Burke-Ernzerhof (PBE) form of the generalized gradient approximation. ABINIT is used to solve the Kohn-Sham (KS) equations, yielding KS states and associated eigenenergies. 
The KS states are used to evaluate the dipole transition matrix elements between the 1s core states to the unoccupied valence states and, with the eigenenergies, are used to construct the density of states (DOS) and the absorption cross-section\cite{humphriesProbingElectronicStructure2020,vinkoDensityFunctionalTheory2014}. 

Molecular dynamics (MD) simulations with SPARC\cite{zhangSPARCV200Spinorbit2024} are carried out to evolve the ionic structure of the disordered WDM system. SPARC is a Kohn-Sham DFT code that provides specialized density-matrix and spectral-partition based methods \cite{suryanarayanaSQDFTSpectralQuadrature2018, sadighSpectralpartitionedKohnShamDensity2023, xuRealspaceDensityKernel2022} to efficiently handle both high-temperature and large-scale calculations. Accurate calculations at such conditions are enabled by the use of hard, localized Optimized Norm Conserving Vanderbilt (ONCV) pseudopotentials\cite{hamannOptimizedNormconservingVanderbilt2013} with multiple projectors in each angular momentum channel to span the required energy space.

The experimental spectrum is measured over the duration of the backlighter and with a much larger sample of ions than possible to use in the simulation, thus a representative average over many respective configurations and many snapshots are needed to properly recreate the system. For the present purpose, SPARC provides an efficient means to evolve the disordered WDM system and gives position information to ABINIT.

The x-ray absorption spectrum at the K-edge is characterized by the coupling of 1s states to the vacant states and the electronic occupation at the edge. The photoabsorption cross-section can be written as\cite{mullerBandstructureApproachXray1984}:

\begin{equation}
  \label{eqn:pixs}
  \sigma(h\nu) = ah\nu\sum_{f}\left|\bra{\psi_{1s}}\mathcal{R}\ket{\psi_{f}}\right|^{2}\delta(E_f - E_{1s} - h\nu)
\end{equation}

\noindent where the constant $a = 4\pi^2\alpha$ (with $\alpha$ being the fine structure constant), $\mathcal{R}$ is the electric dipole moment operator acting on the initial 1s state, $\bra{\psi_{1s}}$, and selecting a p-state among the final states, $\ket{\psi_{f}}$ at a given energy $h\nu$. Equation \ref{eqn:pixs} can be simplified and explicitly written in terms of the energy density of states $\varrho(h\nu)$, the matrix element for transitions, $M$, both of which can obtained from the DFT calculation\cite{humphriesProbingElectronicStructure2020, forteResonantInelasticXray2024}: 

\begin{equation}
  \label{eqn:absorption}
  \sigma(h\nu) \propto \sum_{f}\varrho(h\nu)|M(h\nu)|^{2}[1 - f(h\nu)].
\end{equation}

Here $[1 - f(h\nu)]$ is the electron occupancy, which follows a Fermi-Dirac distribution. This formulation enables the measured absorption spectrum for the warm dense copper plasmas to be compared to a DFT model approach. Both CR and DFT models can reproduce the absorption spectra for ambient copper as compared to reference data. In the CR model case, a global shift of \qty{10}{\electronvolt} was required to align K-shell features (K-edge, K$\alpha$, K$\beta$) to ambient, solid density reference copper measurements. In the DFT model case, absolute energies are not given; rather the model predictions are shown relative to the calculated chemical potential and overlaid to match the experimental K-edge.

The two modeling approaches have complementary strengths and weaknesses. CR models have extensive electronic structure based on detailed electronic configurations that are spectroscopically accurate in the isolated-ion limit. They capture the effects of excitations \cite{akliTemperatureSensitivityCu2007} and charge state distributions by including hundreds of transitions in a single feature: for example, the 1s$\rightarrow$3p transition from Cu$^{5+}$ is several eV lower than the same transition from Cu$^{6+}$, and the same transition from an excited state of Cu$^{5+}$ falls somewhere in between.  

However, CR models treat density effects arising from the local plasma environment as perturbative modifications on the isolated ions using averaged and often ad-hoc models of that environment. Their predictions will depend on the choice of IPD/PPS model, and these corrections are not in general self-consistent. For PPS in particular, both SCRAM and Cretin treat it as a correction in the final calculated output rather than applying the shifts self-consistently during the kinetics solve.

By contrast, DFT models treat electrons as KS states that respond self-consistently to changes in the local environment. Thus, density effects like IPD, PPS, and electron degeneracy emerge naturally as responses to changes in the ionic structure: for example, a 1s$\rightarrow$3p transition in an ion with close neighbors will have a lower energy than the same transition in a more isolated ion. However, even with the most extensive ionic ensembles, each of these KS states represents an averaged electronic configuration and each transition is between fictitious KS orbitals. Thus effects such as orbital relaxation \cite{roseEffectOrbitalRelaxation1986} are excluded and the accuracy of a given DFT model depends on its choices of exchange-correlation functionals and PAW constructions. 

Nevertheless, DFT provides a tractable method for constructing the XAS from the readily calculable components in Equation \ref{eqn:absorption}, and its self-consistent treatment of the local environment is expected to more naturally describe the K-edge dynamics as compared to the CR models. The bound-bound absorption provides an important point of comparison here, as modeling it accurately may be sensitive to detailed excited configurations as treated explicitly in the CR models, as described above.


\section{Comparison with experimental data}\label{sec:comparison}
\subsection{Modeling with collisional-radiative codes}\label{subsec:crmcompare}
The x-ray absorption through the warm dense copper is modeled with Cretin and SCRAM by calculating the opacities over a range of contributing densities and temperatures informed with the simulated hydrodynamic time-evolution of the plasma during the probe window. The temperature and density ranges are informed by simulation, see Table \ref{tab:expvalues}, but temperature is then scaled during the fitting process to the observed bound-bound feature. This method incorporates the temperature gradient but allows for the mean temperature to remain a degree of freedom for the fitting. These opacity contributions are combined to construct the total x-ray transmission. 

Detailed transitions relevant to the K-edge region, such as the 1s$\rightarrow$np transitions, were explicitly added to the FAC data used by Cretin to improve the accuracy of the spectral features near the absorption edge. Both the SP and EK models were used to evaluate the sensitivity of each IPD prescription in this regime, however the EK model was unable to converge to physically realistic solutions and is therefore excluded from this analysis. In this regime, the EK model can produce strong feedback between ionization and IPD, leading to unrealistic spectral predictions, such as the loss of the 3p states. Figure \ref{fig:model_crm}(a) compares the transmission spectra calculated using Cretin with the experimental findings. Cretin is modeled with the density and temperature range method described, at \qty{24}{\gram\per\cubic\centi\metre} and \qty{23}{\electronvolt} as well as at a single condition from the inferred experimental fit density and temperature of \qty{\densmd}{\gram\per\cubic\centi\metre} and \qty[round-mode=figures, round-precision=2]{\tempmd}{\electronvolt}. Similarly, SCRAM with the IS IPD model is modeled at an averaged density and temperature of \qty{24}{\gram\per\cubic\centi\metre} and \qty{25}{\electronvolt}, respectively; in addition to at the experimentally inferred values in Figure \ref{fig:model_crm}(b).

The CR models capture the overall structure and trends of the XAS measurements, most notably the bound-bound absorption's temperature sensitivity between the low- and mid-drive configurations. SCRAM tends to better match the experimentally inferred $\overline{Z}$ and consequently the bound-bound feature position. Broadening from the range of excited states and charge states is present in both models and compares well with the observed line width. Figure \ref{fig:model_crm}(c) compares the modeled CSD's for both conditions compared to the experimental inference, where we see Cretin estimating a larger $\overline{Z}\sim 8.1$ for both and for SCRAM $\overline{Z}\sim 6.9$. Neither Cretin nor SCRAM reproduces the measured K-edge position, with the calculated edge shifts being about 53 eV larger than observed at conditions that are otherwise consistent with the bound-bound feature. 

\begin{figure}
  \includegraphics[width=\linewidth]{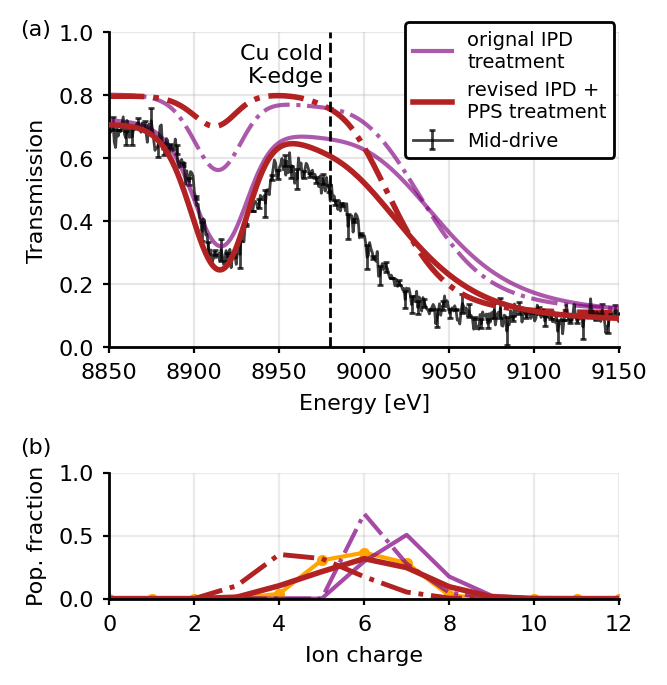}
  \caption{(a) SCRAM with modified PPS treatment and more detailed level inclusion, modeled at two conditions: At averaged conditions  of \qty{24}{\gram\per\cubic\centi\metre} and \qty{30}{\electronvolt} (solid red). Single condition fit (dashed red) at the inferred values from simulation for the density and experiment for the temperature of \qty{\densmd}{\gram\per\cubic\centi\metre} and \qty[round-mode=figures, round-precision=2]{\tempmd}{\electronvolt}, respectively. Comparison with the experimental XAS mid-drive is shown (black) and the previous SCRAM calculations (magenta). (c) Comparison of the CSD of the modified treatment in SCRAM with the previous CR model calculations and experiments.}
  \label{fig:scram_v86}
\end{figure}

In the CR models, the edge position is a combination of three ionization-dependent effects: increases in the 1s binding energy due to decreased screening with ionization, increases in the Fermi energy due to higher electron density, and decreases due to IPD. These effects are not independent, however: increasing the magnitude of the IPD to shift the K-edge to lower energies would lead to more ionization, moving the edge again to higher energies. 

The reasonable agreement in the bound-bound feature at the conditions inferred from the averaged fitting scheme indicates that the atomic data is sufficiently detailed to capture some ion dependence and temperature dependence of the resonance absorption. The discrepancies at the edge indicate that the averaged IPD modifications do not fully capture the electronic behavior at the continuum edge. Similar disagreement with both CR models suggests a broader limitation of the CR model in this regime rather than an issue specific to a single implementation.


The valence structure, and its influence on the photoionization process measured with XAS, requires a self-consistent response of the bound electron wave functions and occupations to changes in free- and bound-electron screening. The absence of a self-consistent treatment also affects the PPS of bound-bound emission and absorption features, which currently relies on an analytical model for hydrogen-like ions. This suggests that the discrepancies are not only due to the magnitude of the IPD or its effect on the ionization balance. They also depend on how IPD and PPS combine to modify the level structure used by the kinetics, which in turn alters the ionization. To explore this sensitivity, we applied a level dependent adjustment of the IPD that includes the associated shifts of PPS. 

This modification predominately alters the M-shell levels for these warm dense copper plasmas. When the PPS correction is included in the level structure, the M shell remains farther from the continuum edge than with IPD alone. This reduces the calculated ionization, shifting the resulting K-edge to lower energy relative to the previous calculations shown in Figure \ref{fig:model_crm}(b). It is also important to ensure that detailed level structure is retained in the lower charge states to enable this reduction in $\overline{Z}$. 

Figure \ref{fig:scram_v86} shows the calculated transmission when considering this modified IPD-PPS implementation with SCRAM. As compared to the previous SCRAM calculation we observe a lower K-edge at the same conditions. The bound-bound features see minimal change, but the added shift to the 3p levels weakens the feature depth as it now sits farther from the continuum, though less affected than the M shell. This modified SCRAM calculation gives $\overline{Z}\sim6.0$, and $T\sim$\qty{29}{\electronvolt} to fit the bound-bound absorption; this results in a reduction of the K-edge discrepancy to approximately \qty{28}{\electronvolt}. For these temperature and density conditions, the calculated chemical potential is approximately \qty{36}{\electronvolt}.

This result suggests that improved CR modeling requires a self-consistent description of the density-dependent atomic structure, that will inherently include IPD and PPS, as well as a smoother description of pressure ionization and state loss to the continuum. Although still approximate, the modified SCRAM treatment moves in this direction with by increasing details of lower charge state ions and direct inclusion of the PPS. These limitations motivate comparison with a DFT-based framework, where the ab initio framework readily includes the effect of density rather than imposed through separate corrections to isolated-ion states.

\subsection{\label{subsec:dftcompare}Modeling with density functional theory}
The DFT approach addresses a central limitation of the CR models by treating density and its effects self-consistently rather than through corrective measures. Here, the x-ray absorption is modeled with SPARC and ABINIT. SPARC computes the ionic motion of the warm dense copper, while ABINIT calculates the eigenenergies, eigenstates, and dipole matrix elements required to construct the x-ray absorption for individual snapshots. 

The MD simulations with SPARC evolve the system from an initial FCC copper lattice to an equilibrated configuration representative of the WDM conditions. The MD stage is used to generate representative disordered WDM configurations, but was not run for the full experimental timescale. ABINIT then calculates the photoabsorption cross-section from equation \ref{eqn:absorption} for selected MD snapshots, and the resulting absorption curves are averaged to construct the final absorption spectrum.

\begin{figure}
  \centering
  \includegraphics[width=\linewidth]{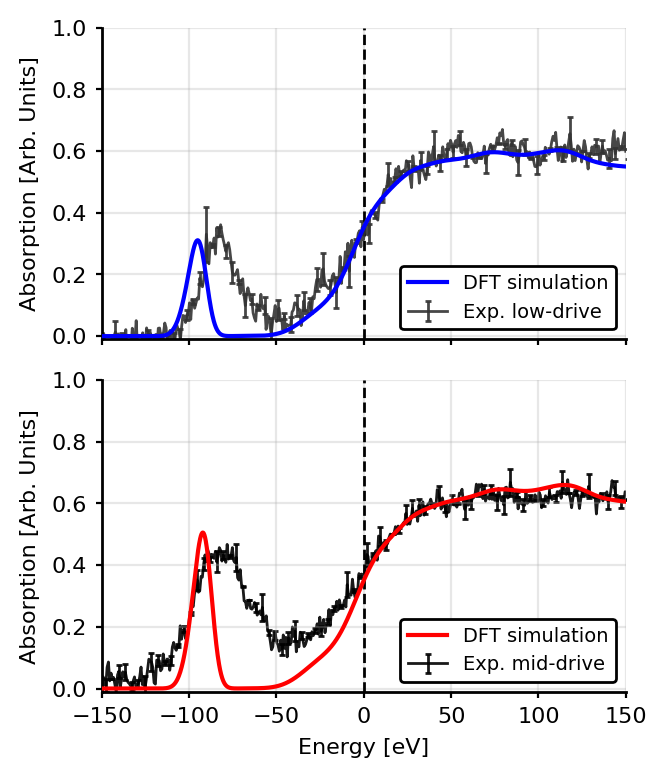}
  \caption{(a) DFT simulation of the x-ray absorption of the low-drive conditions for a 108 atom supercell, at \qty{\dftdens}{\gram\per\cubic\centi\metre} and \qty{\dfttemp}{\electronvolt} compared with experiment. (b) DFT simulation of the x-ray absorption of the mid-drive conditions, at \qty{\dftdensmd}{\gram\per\cubic\centi\metre} and \qty{\dfttempmd}{\electronvolt}.}
  \label{fig:model_dft}
\end{figure}

For WDM conditions, a large plane wave (PW) basis is required to fully characterize the valence electrons in the system. At ambient conditions (\qty{300}{\kelvin}, \qty{8.96}{\gram\per\cubic\centi\metre}) a 108 atom system requires at least a thousand PWs to describe the valence electron structure. At the nearly $3\times$ compression and tens of \unit{\electronvolt} temperatures this quickly can exceed many thousand. The computational cost therefore grows rapidly with system size, limiting the practical scalability, especially in the context of fitting to experimental spectra.

The spectra presented are calculated using a $3$\texttimes$3$\texttimes$3$ unit cell of 108 atoms, which provides a tractable balance between computational expense and accurate representation of the measured XAS. The MD system was evolved using an isokinetic ensemble with Gaussian thermostat for the first picosecond to allow for the pressure to equilibrate. After equilibration, the system was propagated for 12,000 steps with a \qty{0.12}{\femto\second} time step.

In total, the DFT absorption calculations use 60 snapshots spaced approximately \qty{25}{\femto\second} apart. Electronic states are included down to $10^{-4}$ which necessitates 5500 bands for these calculations. Brillouin zone sampling is done at the $\Gamma$ point, which is expected to be a good approximation for the disordered WDM copper cell. We used a PAW data set generated from ATOMPAW with 19 valence electrons ($3s^{2}3p^{6}3d^{10}4s^{1}$) and cutoff radius of \qty{1.2}{\bohr} and a plane wave cutoff of \qty{45}{\hartree}.

The natural linewidth of the levels involved in the absorption process contribute to the spectral broadening around the edge. A Lorentzian function model was used to represent the delta function, from Equation \ref{eqn:pixs}, following Ref.\cite{jourdainUnderstandingXANESSpectra2020}:

\begin{equation}
  \label{eqn:deltabroad}
  \delta(E_f - E_{1s} - h\nu) = \frac{1}{\pi}\frac{\Gamma_f(h\nu)}{\Gamma_f(h\nu)^2 + (E_f - E_{1s} - h\nu)^2}
\end{equation}

This Lorentzian function has the energy-dependent width $\Gamma_f$,

\begin{equation}
  \label{eqn:gammaF}
  \Gamma_f = \begin{cases}
    0 & h\nu < E_{Fermi} \\
    \Gamma_{hole} + \gamma(\nu) & h\nu > E_{Fermi}
  \end{cases}
\end{equation}

$\Gamma_{hole}$ is taken to be the tabulated K level width of \qty{1.55}{\electronvolt}\cite{krauseNaturalWidthsAtomic1979} and $\gamma(\nu)$ is an empirical function following Ref.\cite{bunauProjectorAugmentedWave2013} with a maximum \qty{6.1}{\electronvolt}. An additional Gaussian convolution accounts for the experimental instrument response and is chosen to match the edge width of the cold experiment XAS with FWHM of \qty{7}{\electronvolt}.

Figure \ref{fig:model_dft}(a) compares the low-drive XAS measurement with the DFT calculation at \qty{\dftdens}{\gram\per\cubic\centi\metre} and \qty{\dfttemp}{\electronvolt}. At this density, the chemical potential is approximately \qty{34}{\electronvolt}. The experimental data has been inverted to absorption and shifted so that zero energy corresponds to the bottom of the experimental K-edge. The DFT calculation is aligned to bottom of the valence band. This relative alignment prioritizes comparing the model's ability to reproduce the spacing and shape of the K-edge absorption features.

For this comparison, we find improved agreement between experiment and simulation. The relative position of the 1s$\rightarrow$3p contribution to the K-edge agrees more closely with the experiments than with the CR model calculation. The separation, however, remains \qty{10}{\electronvolt} and \qty{14}{\electronvolt} wider than measured for mid- and low-drive, respectively. This result suggests that the self-consistent treatment of the electronic structure captures the density driven modifications of the valence states that control the resulting K-edge position.

The comparison also reveals other areas of disagreement; the calculation lacks an accurate absolute energy scale needed for forward fitting directly to the measured spectra and does not include the effects from excited state and charge state distributions. Both excited state configurations and charge state distribution broaden and shift the observed absorption features. The calculated 3p binding remains slightly offset from the measurement. The position of this feature is only weakly sensitive over the density range considered here. However, the p-like DOS at near the continuum shows a stronger sensitivity to degeneracy, and therefore density, and the best agreement is found at a higher density than predicted by the hydrodynamic simulations. This may reflect the limitations of the equation-of-state treatment used in the hydrodynamic code under these conditions, although the DFT result remains within the uncertainties of the density calculation from HYDRA. 

The single condition calculation captures the averaged K-edge structure but does not fully include the temporal gradients in the experimental system. The combined lack of explicit excited configurations, a CSD, and temporal gradient is most evident in the comparison with the mid-drive at \qty{\dfttempmd}{\electronvolt}. DFT captures the overall features of the K-edge structure, but as can be seen in Figure \ref{fig:model_dft}(b), the simplifications in this calculation lead to disagreement in the 1s$\rightarrow$3p feature. DFT also has important limitations including sensitivity to the choice of exchange-correlation functional, and PAW construction at these high densities, as well as finite effects from the calculation supercell size.

Overall, the DFT spectra capture the broad behavior of the K-edge of warm dense copper more successfully than the CR prescription, particularly in the K-edge itself, where the ad-hoc IPD models cannot reproduce the complex valence and continuum electron structure. The remaining discrepancies highlight some limitations of DFT in this domain but nevertheless DFT provides a powerful framework for modeling in the WDM regime with the appropriate PAW potentials and multistep averaging. These results motivate the need for improved models that can combine the efficiency and charge-state resolution of CR methods into the density-dependent electronic-structure information like that from DFT.

\section{Conclusions}\label{sec:conclusion}
In summary, we have presented improved x-ray absorption measurements of warm dense copper using a buried layer planar platform at the OMEGA laser facility. The experimental platform builds on a series of experimental iterations while incorporating an improved target design, shielding, and backlighter improvements. These changes improved the spectral resolution of the copper K-edge and further reduced the influence of spatial and temporal gradients during the measurement.  

The temperature and charge state distribution was inferred from a functional fit of the K-edge slope and 1s$\rightarrow$3p resonance absorption. Densities are estimated from HYDRA hydrodynamic simulations previously adjusted with the aid of experimental VISAR measurements. The two experimental spectra show sensitivity of the K-edge to temperature at densities of approximately $\densmd\pm 3$ \unit{\gram\per\cubic\centi\metre}, while the mid-drive configuration shows increased thermal broadening from higher achieved temperatures. The CR and DFT calculations that provide the best agreement with measured are consistent, within uncertainty, of the HYDRA inferred density, but favor higher temperatures of \qtyrange[range-units=single]{22}{29}{\electronvolt} than that of the Fermi-Dirac model fit. These conditions necessitated the need for a full model of the absorption spectrum rather beyond the simple inference of temperature from the edge slope. 

CR calculations using Cretin and FAC offer several advantages when modeling these spectra including detailed inclusion of emission and absorption lines for a distribution of charge states. Improvements to the Cretin's input atomic data and the inclusion of PPS were made, but we found that disagreements with the measured spectra remained. SCRAM showed a significant improvement with the consideration of more detailed level information for the involved charge states and with the incorporation of PPS into the level structure and kinetics calculation. 

As a result of this inclusion, the modeled ionization is slightly reduced and the K-edge and bound-bound features lie closer together as observed in the measurement, though a \qty{28}{\electronvolt} separation discrepancy remained. This highlights that a consistent treatment of density effects is a necessary component for accurately determining the plasma behavior in this regime.

Neither SP, EK, nor IS IPD models can reproduce the observed behavior of the K-edge position. This discrepancy highlights the challenge in representing the complex effects in a dense plasma through an averaged ad-hoc correction to the isolated atom. In particular, the balance of continuum lowering, ionization, and degeneracy have a strong influence on the observed spectra.

DFT calculations provided an ab initio framework to treat density effects self-consistently. Agreement with a DFT approach underscores the importance of a self-consistent quantum mechanical model to characterize the x-ray absorption in the WDM regime. The DFT modeling at these relatively hot and dense conditions required careful consideration of the potentials and calculation methods, extending this method using direct calculation of the dipole matrix elements to more extreme conditions. DFT also has its limitations, including sensitivity to the exchange-correlation functional and pseudopotential treatments, and the absence of broadening effects. In spite of these limitations, the computed x-ray absorption edge exhibits the correct K-edge behavior and can be used for constraint and inference of plasma conditions. 

Together, these measurements provide a useful benchmark for testing spectral models in WDM and further experiments are needed to characterize the regime. Future theoretical work will tackle a deeper self-consistent integration of density in the CR models to better model spectra in the WDM regime. An important next step for DFT is the integration of more complex line-broadening effects, namely from the multiple charge states in these plasmas.

\section{Acknowledgements}\label{sec:acknowledge}
This work was performed under the auspices of the U.S. Department of Energy by Lawrence Livermore National Laboratory under Contract DE-AC52-07NA27344 and was supported by the LLNL-LORD Program under Project No. 22-ERD-005. SNL is managed and operated by NTESS under DOE NNSA contract DE-NA0003525. This work is partially supported by the Department of Energy, National Nuclear Security Administration under Award No. DE-NA0004147. This material is based upon work supported by the Department of Energy [National Nuclear Security Administration] University of Rochester “National Inertial Confinement Fusion Program” under Award Number(s) DE-NA0004144.

\bibliography{references.bib}
\end{document}